\documentclass[aps,prd,nofootinbib,twocolumn,superscriptaddress,preprintnumbers,balancelastpage,longbibliography]{revtex4-2}

\usepackage{placeins}
\usepackage{amsmath,amssymb,mathtools,bm}
\usepackage{graphicx, color, hepunits}
\usepackage[dvipsnames]{xcolor}
\usepackage{cancel}
\usepackage[normalem]{ulem}
\usepackage{float}
\usepackage{filecontents}
\usepackage{multirow}
\usepackage{mhchem}
\usepackage{tikz}
\usepackage{subfigure}
\usepackage{ragged2e}

\usetikzlibrary{decorations.pathmorphing}
\usetikzlibrary{arrows}
\usepackage{circuitikz}
\usetikzlibrary{arrows.meta}
\usetikzlibrary{shapes.misc}
\usetikzlibrary{positioning}
\usetikzlibrary{decorations.markings}

\usepackage[compat=1.1.0]{tikz-feynman}

\usepackage{hyperref} 
\hypersetup{
    colorlinks=true,      
    linkcolor=blue,        
    citecolor=blue,        
    filecolor=magenta,     
    urlcolor=blue
}
\usepackage[utf8]{inputenc}
\usepackage[english]{babel}

\usetikzlibrary{shapes.misc}
\tikzset{cross/.style={cross out, draw=black, minimum size=2*(#1-\pgflinewidth), inner sep=0pt, outer sep=0pt}, 
cross/.default={1pt}}

\begin{document}

{\flushright KEK-TH-2847
\\[-9ex]
}

\title{Searches for GeV-Scale ALPs at RHIC}

\author{Kaori Fuyuto}
\email{kfuyuto@post.kek.jp}
\affiliation{KEK Theory Center, IPNS, KEK, Tsukuba 305–0801, Japan}
\affiliation{RIKEN-Berkeley Center, RIKEN iTHEMS, Wako, Saitama 351-0198, Japan}

\author{Claudio Andrea Manzari}
\email{manzari@ias.edu}
\affiliation{School of Natural Sciences, Institute for Advanced Study, Princeton, NJ, 08540, USA}

\author{Hitoshi Murayama}
\email{hitoshi@berkeley.edu}
\affiliation{Leinweber Institute for Theoretical Physics, University of California, Berkeley, CA 94720, U.S.A.}
\affiliation{Kavli Institute for the Physics and Mathematics of the Universe (WPI), University of Tokyo, Kashiwa 277-8583, Japan}
\affiliation{Theoretical Physics Group, Lawrence Berkeley National Laboratory, Berkeley, CA 94720, U.S.A.}

\begin{abstract}
We point out that ultra-peripheral Au+Au collision data collected at the Relativistic Heavy Ion Collider, operational during 2000-2026, can be used to search for axion-like particles coupled to photons via the resonant process $\gamma\gamma \to a \to \gamma\gamma$. Exploiting the $Z^4$ enhancement of the two-photon luminosity in heavy-ion collisions and the low photon energy thresholds achievable at RHIC, we simulate signal and background processes, the latter dominated by light-by-light scattering, hadronic resonance production, and misidentified $e^+e^-$ pairs, and estimate upper limits on the ALP-photon coupling $g_{a\gamma\gamma}$ assuming $1.9~\text{nb}^{-1}$ of existing data collected by the PHENIX experiment. We find sensitivity to ALP masses in the range $2~\text{GeV} \lesssim m_a \lesssim 5~\text{GeV}$ with couplings $g_{a\gamma\gamma} \gtrsim 4\times 10^{-4}~\text{GeV}^{-1}$, probing previously unexplored regions of parameter space. Access to larger luminosity datasets could substantially extend the sensitivity of this search, motivating a dedicated analysis of ultra-peripheral collision data collected at RHIC by PHENIX as well as other experiments.
\end{abstract}
\maketitle

\section{Introduction}
The Standard Model (SM) of particle physics has successfully passed virtually every empirical test in the laboratory so far. However, our Universe still holds several puzzles that cannot be explained by the SM, such as the strong CP problem, the flavor puzzle, the baryon asymmetry of the Universe and the nature of dark matter. Among the most well-motivated extensions of the SM addressing these puzzles are the QCD axion and the broader class of \textit{axion-like particles} (ALPs). ALPs are pseudo-scalar fields that arise as pseudo-Nambu-Goldstone bosons associated with the spontaneous breaking of global symmetries. Beyond their original role as a solution to the strong CP problem, ALPs have been extensively studied in a wide range of contexts: as dark matter candidates~\cite{Preskill:1982cy,Abbott:1982af,Dine:1982ah}, in connection with baryogenesis \cite{Kuzmin:1992up, Servant:2014bla}, in explanations of the observed hierarchies of fermion masses and mixings \cite{Froggatt:1978nt, Ema:2016ops, Calibbi:2016hwq}, within Grand Unified Theories~\cite{Agrawal:2022lsp}, models with extra spatial dimensions~\cite{deGiorgi:2024elx}, and scenarios of cosmic inflation~\cite{Choi:2014rja}. Moreover, ALPs are now understood to be a generic prediction of string theory compactifications~\cite{Svrcek:2006yi,Arvanitaki:2009fg,Cicoli:2012sz,Demirtas:2018akl,Halverson:2019cmy,Demirtas:2021gsq}, motivating the so-called \textit{axiverse}. In this framework, the QCD axion is accompanied by a rich spectrum of additional axion-like states, often characterized by an approximately log-flat mass distribution that naturally encompasses the range relevant for this work~\cite{Green:1984sg,Witten:1984dg,Svrcek:2006yi,Arvanitaki:2009fg,Halverson:2019cmy,Conlon:2006tq,Acharya:2010zx,Ringwald:2012cu,Cicoli:2012sz,Demirtas:2021gsq,Mehta:2021pwf,Gendler:2023kjt}.\\

Depending on their mass and couplings, ALPs are actively searched for using a combination of astrophysical and cosmological probes, as well as laboratory-based experiments. In particular, for the ALPs in the 5--100 GeV mass range, light-by-light scattering in ultra-peripheral heavy-ion collisions (UPCs) at the Large Hadron Collider (LHC) provides one of the most sensitive search channels \cite{Knapen:2016moh, ATLAS:2020hii, CMS:2024bnt}. These searches target the resonant production of an ALP in the diphoton final state, $\gamma\gamma \to a \to \gamma\gamma$, which appears as a narrow resonance in light-by-light scattering events. The sensitivity of these searches is enhanced by a factor of $Z^4$ relative to proton-proton collisions. The same strategy can be extended to lower collision energies. In this context, the Relativistic Heavy Ion Collider (RHIC) at Brookhaven National Laboratory is particularly well suited for such searches. Operating from 2000 to 2026, RHIC collided nuclear species ranging from protons to gold nuclei at center-of-mass energies up to $\sqrt{s}=200$ GeV, pursuing a broad program in nuclear and particle physics. Alongside its flagship studies, RHIC accumulated substantial datasets of ultra-peripheral heavy-ion collisions, providing a unique laboratory for precision studies of photon-induced reactions and searches for physics beyond the SM.
 
At the LHC, the high energies, pileup and trigger limitations require typical photon energy thresholds of $\sim 2.5$ GeV, and therefore sensitivities to ALPs masses above $\sim 5$ GeV~\cite{ATLAS:2020hii, CMS:2024bnt}.
In contrast, we point out that experiments at RHIC operated at a lower collision energy and with a relatively low instantaneous luminosity. These conditions allow photons to be reconstructed and reliably identified down to lower energies, allowing to probe ALPs in a lower-mass region of parameter space that is not accessible in current LHC heavy-ion searches. Such relatively light ALP mass region will also be accessible at current and future experiments such as  ALICE \cite{dEnterria:2022sut, Khatun:2024vgn}, Belle II~\cite{Dolan:2017osp}, and the EIC \cite{Balkin:2023gya}.

The RHIC experimental program included four main detectors, PHENIX, STAR, BRAHMS, and PHOBOS, all of which collected Au+Au collision data. Among these, PHENIX and STAR accumulated the largest integrated luminosities and were equipped with electromagnetic calorimeters capable of reconstructing photons, making them the most promising candidates for a UPC-based ALP search. More recently, sPHENIX, a major rebuild of the PHENIX detector began operations in 2023 and collected Au+Au data in the final RHIC runs. 

Whether dedicated UPC datasets with the appropriate trigger configuration and sufficient integrated luminosity were recorded and preserved by STAR, sPHENIX, or other experiments remains to be established, and a systematic survey of the available data would be needed to assess their potential for this search. To the best of our knowledge, neither STAR nor sPHENIX operated a dedicated UPC trigger suitable for this search during their respective runs\footnote{Private communications with members of the STAR and sPHENIX collaborations.}. In the case of STAR, all available electromagnetic calorimeter (EMC) based triggers required additional detector activity beyond back-to-back electromagnetic clusters, such as a reconstructed vertex or hits in the time-of-flight system, making them unsuitable for the selection of exclusive two-photon events. For sPHENIX, EMC-based triggers were available for $p+p$ collisions but not for Au+Au; towards the end of the final run, a small amount of Au+Au data was collected with an EMC trigger, though the integrated luminosity is small and the trigger threshold of $p_T \gtrsim 3~\text{GeV}$ would in any case limit sensitivity to ALP masses already constrained by LHC measurements. 

In this work, we focus on the PHENIX experiment, for which a UPC trigger was operational and an integrated luminosity of $1.9~\text{nb}^{-1}$ of Au+Au collisions at $\sqrt{s_{NN}} = 200~\text{GeV}$ has been reported~\cite{PHENIX:2009xtn, Takahara:2012upc}, providing a well-characterized reference luminosity on which to base our projected sensitivity estimates.\\

Concretely, we derive projected exclusion limits on the ALP-photon coupling by studying the interaction described by the following Lagrangian
\begin{equation}
\mathcal{L}_{a\gamma\gamma} = \frac{1}{2}\partial_{\mu}a\partial^{\mu}a + \frac{1}{2}m_a^2\,a^2 -\frac{1}{4}\,g_{a\gamma\gamma}\,a\, F_{\mu\nu} \tilde{F}^{\mu\nu},
\label{eq:Lagrangian}
\end{equation}
where $F_{\mu\nu}$ is the electromagnetic field strength tensor, $\tilde{F}^{\mu\nu} = \frac{1}{2}\epsilon^{\mu\nu\rho\sigma}F_{\rho\sigma} $ its dual, and $g_{a\gamma\gamma}$ denotes the ALP--photon coupling. This interaction mediates the production of an ALP through photon-photon fusion as well as its decay into a pair of photons. As a result, ALPs constitute a well-defined experimental target at colliders and, in particular, in UPCs as discussed in the next section. We find that the PHENIX experiment is sensitive to a broad range of ALP masses and couplings, and can probe previously unexplored regions of heavy-axion parameter space for masses of order a few GeV.

\section{Photon--photon interaction}

UPCs occur when two relativistic ions pass each other with an impact parameter larger than the sum of their nuclear radii. In this regime, hadronic interactions are strongly suppressed, and the ions interact predominantly through their electromagnetic fields. For relativistic ions with $Z\gg 1$, the electric and magnetic fields are perpendicular, and the configuration can be represented as a flux of almost-real photons following the Weizsäcker-Williams approximation~\cite{vonWeizsacker:1934nji,Williams:1934ad}. For a relativistic ion with Lorentz boost factor $\gamma \gg 1$, the equivalent photon spectrum can be expressed in terms of the photon energy $k$ and the transverse distance $\bm b$ from the ion trajectory. The number of photons emitted per unit energy and transverse area is given by~\cite{Klein:2016yzr}
\begin{align}
    \frac{dN_\gamma(k,\bm{b})}{dk \, d^2b}
    = \frac{Z^2\alpha}{\pi^2}\,
    \frac{k}{\gamma^2}\,
    K_1^2\!\left(\frac{k|\bm b|}{\gamma}\right)\,,
    \label{eq:PhotonDensity}
\end{align}
where $Z$ is the ion charge, $\alpha$ is the electromagnetic fine-structure constant, and $K_1$ denotes a modified Bessel function of the second kind. Since the exchanged photons are quasi-real, only the contribution from transversely polarized photons is included in Eq.~\eqref{eq:PhotonDensity}.

In an ultra-peripheral collision, the differential two-photon luminosity is obtained by folding the photon fluxes of the two ions,
\begin{align}
\begin{split}
    \frac{d^2N_{\gamma\gamma}(k_1,k_2)}{dk_1\,dk_2}
    = {} & \int d^2b_1\, d^2b_2 \,
    P_{\rm NOHAD}\!\left(|\bm{b}_1-\bm{b}_2|\right) \\
    & \times
    \frac{dN_\gamma(k_1,\bm{b}_1)}{dk_1 \, d^2b_1}\,
    \frac{dN_\gamma(k_2,\bm{b}_2)}{dk_2 \, d^2b_2}\,,
\end{split}
\end{align}
which exhibits the characteristic $Z^4$ enhancement of photon-induced processes in heavy-ion collisions. The suppression of hadronic interactions is enforced by the factor $P_{\rm NOHAD}(b)$, which represents the probability that no hadronic interaction occurs at an impact parameter $b$. Following the implementation in the \texttt{STARlight} code~\cite{Klein:2016yzr}, this probability is modeled as
\begin{align}
    P_{\rm NOHAD}(b) = e^{-\sigma_{\rm NN}\,T_{\rm AA}(b)}\,,
\end{align}
where $\sigma_{\rm NN}$ is the inelastic nucleon--nucleon cross section and
\begin{equation}
T_{\rm AA}(b) = \int d^2b_1\, T_A(b_1)\, T_A(|\bm{b}_1 - \bm{b}|)\,,
\end{equation}
is the nuclear overlap function, constructed from the Woods--Saxon transverse density distributions $T_A(b)$ of the two nuclei.

We focus on photon-photon fusion processes producing a single axion with invariant mass $W = \sqrt{4 k_1 k_2}$ and rapidity
$Y = \tfrac{1}{2}\ln(k_1/k_2)$. In this case, the total cross section factorizes into the two-photon luminosity associated with the colliding ions and the subprocess cross section for axion production in photon-photon collisions,
\begin{align}
    \sigma(\ce{Au}\,\ce{Au} \to \ce{Au}\,\ce{Au}\, a)
    = \int dY\, dW \,
    \frac{d^2N_{\gamma\gamma}}{dW\, dY}\,
    \sigma_{\gamma\gamma}(W)\,.
\end{align}
The cross section for producing a single axion in a $\gamma\gamma$ interaction is given by
\begin{align}
\begin{split}
    \sigma_{\gamma\gamma}(W)
    &= 4\pi \,
    \frac{\Gamma_{\gamma\gamma}\,\Gamma}
    {(W^2 - m_a^2)^2 + m_a^2 \Gamma^2} \\
    &\simeq
    4\pi^2 \,
    \frac{\Gamma_{\gamma\gamma}}{2 m_a^2}\,
    \delta(W - m_a)\,,
\end{split}
\label{sigmagg}
\end{align}
where $m_a$ denotes the axion mass and $\Gamma$ its total decay width. In the second line we have applied the narrow-width approximation, valid for $\Gamma \ll m_a$.
The partial decay width into two photons is
\begin{align}
    \Gamma_{\gamma\gamma}
    = \frac{g_{a\gamma\gamma}^2}{64\pi}\,m_a^3\,.
\end{align}
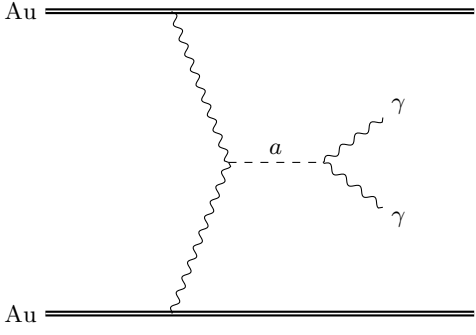
\begin{figure}
\centering
\begin{tikzpicture}
\begin{feynman}

\vertex (N1L) at (-3,2) {Au};
\vertex (N1R) at (3,2);
\vertex (N2L) at (-3,-2) {Au};
\vertex (N2R) at (3,-2);

\vertex (v1) at (-1,2);
\vertex (v2) at (-1,-2);

\vertex (v1bis) at (-0.25,0.);
\vertex (v2bis) at (1,-0);

\vertex (e1) at (2,0.75) {$\gamma$};
\vertex (e2) at (2,-0.75) {$\gamma$};

\diagram*{
    (N1L) -- [double, thick] (N1R),
    (N2L) -- [double, thick] (N2R),

    (v1) -- [photon] (v1bis),
    (v2) -- [photon] (v1bis),

    (v1bis) -- [scalar, edge label=$a$] (v2bis),
    (v2bis) -- [photon] (e1),
    (v2bis) -- [photon] (e2)
        };
\end{feynman}
\end{tikzpicture}
\caption{Feynman diagrams for an ALP production and subsequent decay in two photons in Au-Au UPCs.}
\label{fig:signal_diagram}
\end{figure}

\begin{figure}
    \centering
    \includegraphics[width=1.0\linewidth]{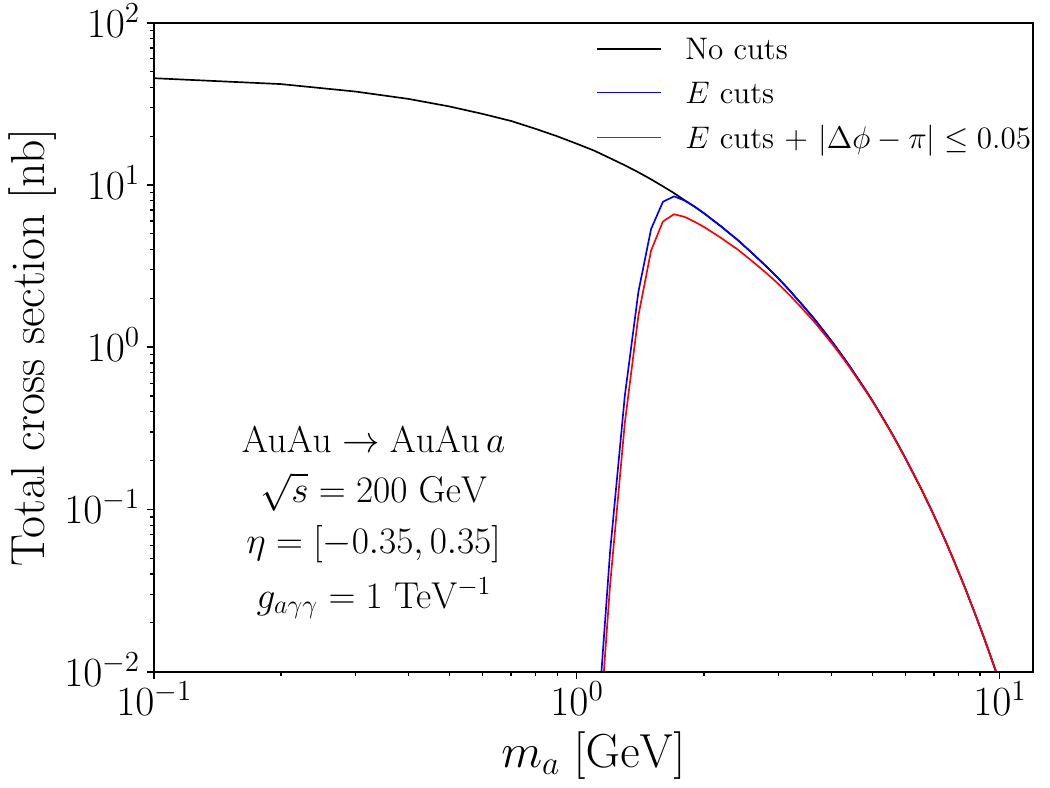}
    \caption{Total cross section $\sigma(\ce{Au}\,\ce{Au} \to \ce{Au}\,\ce{Au}\, a) $ against the axion mass $m_a$ with (blue and red) and without (black) kinematic cuts.}
    \label{fig:signal}
\end{figure}

\begin{figure}
    \centering
    \includegraphics[width=1.0\linewidth]{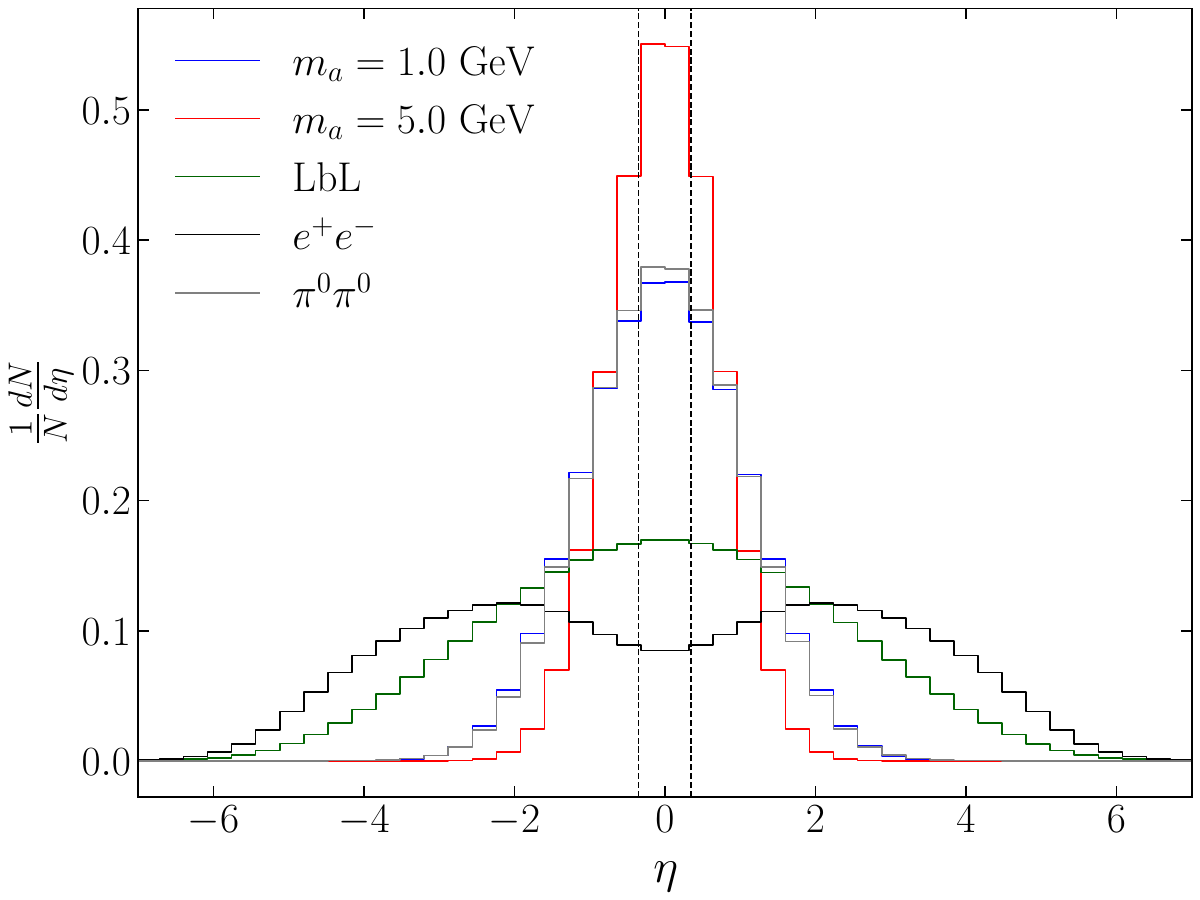}
    \caption{Pseudorapidity distribution of the final-state photons for two values of the ALP mass, compared to the LbL scattering background with $0.1~\mathrm{GeV} \leq \sqrt{s} \leq 10~\mathrm{GeV}$. The smearing of the photon momenta due to detector resolution has not been included. The vertical dashed line show the geometrical acceptance of the PHENIX EM calorimeter.}
    \label{fig:eta_distr}
\end{figure}

\section{Analysis \& Background}

Having described ALPs production in photon–photon interactions, we now consider its experimental signature via its decay products. We focus on the scenario BR($a\rightarrow \gamma\gamma)=1$, in which the ALP decays exclusively to two photons, yielding a resonant diphoton signal in the final state, as shown in Fig.~\ref{fig:signal_diagram}. For the analysis of the data collected by the PHENIX experiment, we assume the trigger for UPC in Au-Au collisions described in Ref.~\cite{PHENIX:2009xtn}, which requires:
\begin{itemize}
\item a veto on coincident signals in both Beam-Beam Counters (covering $3.0 <|\eta| <  3.9$ and full azimuth);
\item the electro-magnetic calorimeter (EMCal), whose geometrical acceptance is $|\eta|\leq 0.35$ and full azimuth, with a $2\times 2$ tile threshold at 0.8 GeV;
\item and at least 30 GeV energy deposited in one or both of the zero degree calorimeters (ZDCs) to
select Au+Au events with forward neutron emission.
\end{itemize}
The integrated luminosity of Au+Au collisions collected in the 2004, 2007 and 2010 runs, using this trigger, amounts to $\mathcal{L}\simeq 1.9\; {\rm nb}^{-1}$~\cite{PHENIX:2009xtn, Takahara:2012upc}. We further require the detection of exactly two photons with an azimuthal opening angle $|\Delta\phi - \pi|\leq 0.05$, to select back-to-back photons in the transverse plane, and assume a threshold sensitivity of $E_{\gamma} = 0.3$ GeV, and an energy resolution of $\sigma_E / E \simeq 8\% \sqrt{\mathrm{GeV}/E}$ as reported in Ref.~\cite{PHENIX:2003fvo}.\\

We used the public code \texttt{STARlight}~\cite{Klein:2016yzr} to generate signal and background events, requiring Coulomb break up of one or both nuclei, with no restriction on the number of
neutrons emitted. For simplicity, all neutrons emitted in these dissociation processes were assumed to fall within the acceptance of the ZDCs. Scanning the ALP mass range
$0.1\,\mathrm{GeV} \leq m_a \leq 10\,\mathrm{GeV}$, we generated $10^6$ event for each mass point. The resulting production cross section, together with the impact of our selection cuts on $E_{\gamma}$, $\Delta\phi$, and $\eta$, is shown in Fig.~\ref{fig:signal}. Due to the imposed $E_{\gamma}$ requirements, our analysis is sensitive only to ALPs with $m_a \gtrsim 1\,\mathrm{GeV}$. In this mass range, for $g_{a\gamma\gamma} = 1\, {\rm TeV}^{-1}$, the cross section is of order $\sim 10\,\mathrm{nb}$ at $m_a \simeq 2\,\mathrm{GeV}$, decreasing to $\sim 0.01\,\mathrm{nb}$ at $m_a = 10\,\mathrm{GeV}$. Consequently, for an integrated luminosity of $1.9\,\mathrm{nb}^{-1}$, we expect between $\mathcal{O}(20)$ and $\mathcal{O}(0.1)$ signal events for $1\,\mathrm{GeV} \leq m_a \leq 10\,\mathrm{GeV}$.

\begin{figure}
    \centering
    \includegraphics[width=1.0\linewidth]{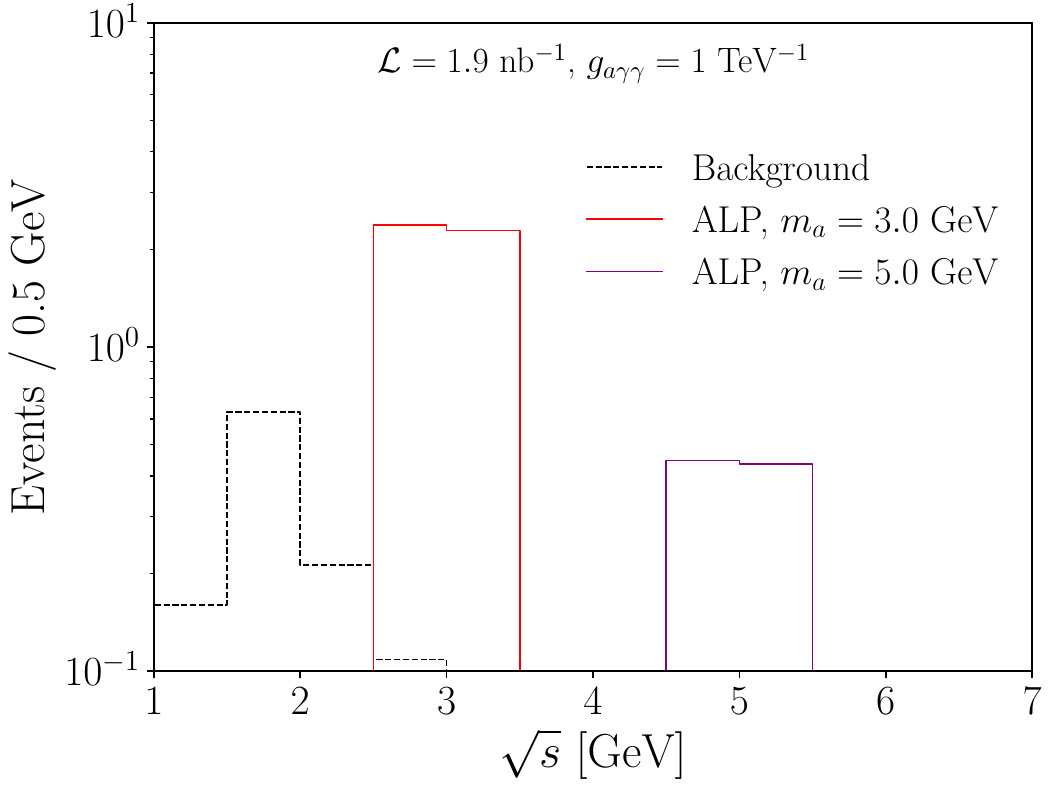}
    \caption{Expected number of events as a function of the di-photon invariant mass for a total integrated luminosity of $1.9~\mathrm{nb}^{-1}$. The background is dominated by the production of $\eta^{\prime}$, $\eta_c$, LbL scattering and $e^+e^-$ pair production. The signal expected for $g_{a\gamma\gamma} = 1\; {\rm TeV}^{-1}$ and two illustrative values of $m_a$ is also shown. The data are binned using a uniform bin width of $0.5\,\mathrm{GeV}$. BR($a\rightarrow \gamma\gamma)=1$ is assumed.}
    \label{fig:events}
\end{figure}

In the region of parameter space relevant for our study, we find $\Gamma_a / m_a \lesssim 10^{-2}$, which justifies the use of the narrow-width approximation.
As a result, the invariant mass distribution of the final-state photons can be effectively treated as a Dirac $\delta$-function centered at $m_a$.
The corresponding pseudorapidity distributions are shown in Fig.~\ref{fig:eta_distr}.
For benchmark masses $m_a = 5\,\mathrm{GeV}$ and $m_a = 1\,\mathrm{GeV}$, the fraction of events falling within the electromagnetic calorimeter acceptance is approximately $35\%$ and $24\%$, respectively. The detector energy resolution is modeled by smearing the photon energy according to the EMC resolution,
$\sigma_E / E \simeq 8\% \sqrt{\mathrm{GeV}/E}$~\cite{PHENIX:2003fvo}.
For each photon, the true energy is smeared with a Gaussian of width $\sigma_E = 0.08 \sqrt{E/{\rm GeV}}$ GeV, while the momentum direction is kept fixed.
The reconstructed four-momentum is then obtained by rescaling the spatial components accordingly.
This procedure is applied on an event-by-event basis and propagated consistently through the analysis. The number of events expected for the two benchmark cases $m_a = 3$ GeV and $m_a = 5$ GeV and $g_{a\gamma\gamma} = 1\, {\rm TeV}^{-1}$ is shown in Fig.~\ref{fig:events}.
\\

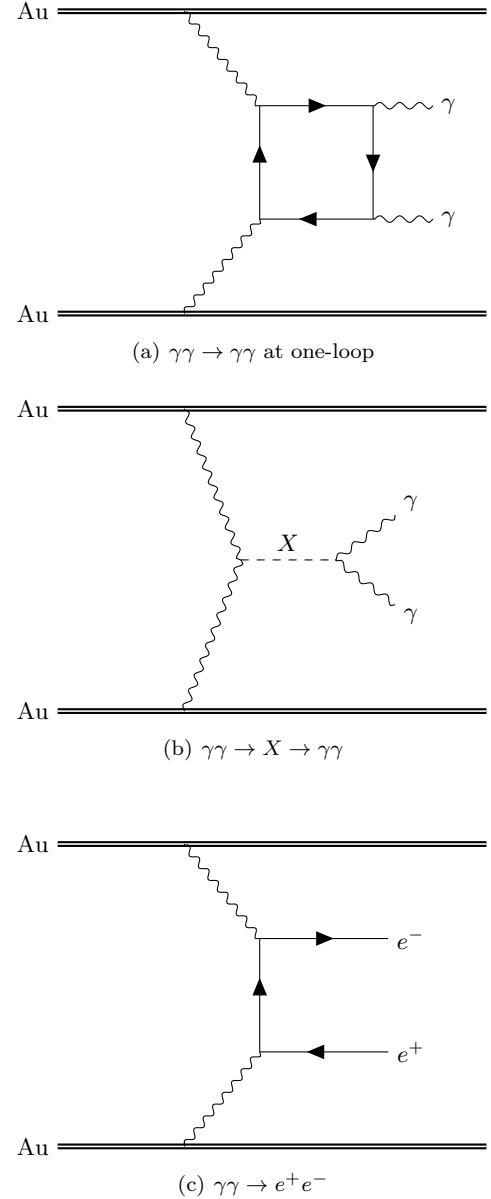
\begin{figure}
\centering

\subfigure[$\gamma\gamma \to \gamma\gamma$ at one-loop\label{fig:lbl}]{%
\begin{tikzpicture}
\begin{feynman}
    \vertex (N1L) at (-3,2) {Au};
    \vertex (N1R) at (3,2);
    \vertex (N2L) at (-3,-2) {Au};
    \vertex (N2R) at (3,-2);

    \vertex (v1) at (-1,2);
    \vertex (v2) at (-1,-2);

    \vertex (v1bis) at (0,0.75);
    \vertex (v2bis) at (0,-0.75);
    \vertex (e1) at (1.5,0.75);
    \vertex (e2) at (1.5,-0.75);

    \vertex (o1) at (2.5,0.75) {$\gamma$};
    \vertex (o2) at (2.5,-0.75) {$\gamma$};

    \diagram*{
        (N1L) -- [double, thick] (N1R),
        (N2L) -- [double, thick] (N2R),
        (v1) -- [photon] (v1bis),
        (v2) -- [photon] (v2bis),
        (v1bis) -- [fermion] (e1),
        (v1bis) -- [anti fermion] (v2bis),
        (v2bis) -- [anti fermion] (e2),
        (e1) -- [fermion] (e2),
        (e1) -- [photon] (o1),
        (e2) -- [photon] (o2)
    };
\end{feynman}
\end{tikzpicture}
}
\hfill
\subfigure[$\gamma\gamma \to X \to \gamma\gamma$\label{fig:res}]{%
\begin{tikzpicture}
\begin{feynman}
    \vertex (N1L) at (-3,2) {Au};
    \vertex (N1R) at (3,2);
    \vertex (N2L) at (-3,-2) {Au};
    \vertex (N2R) at (3,-2);

    \vertex (v1) at (-1,2);
    \vertex (v2) at (-1,-2);

    \vertex (v1bis) at (-0.25,0.);
    \vertex (v2bis) at (1,0);

    \vertex (e1) at (2,0.75) {$\gamma$};
    \vertex (e2) at (2,-0.75) {$\gamma$};

    \diagram*{
        (N1L) -- [double, thick] (N1R),
        (N2L) -- [double, thick] (N2R),
        (v1) -- [photon] (v1bis),
        (v2) -- [photon] (v1bis),
        (v1bis) -- [scalar, edge label=$X$] (v2bis),
        (v2bis) -- [photon] (e1),
        (v2bis) -- [photon] (e2)
    };
\end{feynman}
\end{tikzpicture}
}

\vspace{0.5cm}

\subfigure[$\gamma\gamma \to e^+ e^-$\label{fig:ee}]{%
\begin{tikzpicture}
\begin{feynman}
    \vertex (N1L) at (-3,2) {Au};
    \vertex (N1R) at (3,2);
    \vertex (N2L) at (-3,-2) {Au};
    \vertex (N2R) at (3,-2);

    \vertex (v1) at (-1,2);
    \vertex (v2) at (-1,-2);

    \vertex (v1bis) at (0,0.75);
    \vertex (v2bis) at (0,-0.75);

    \vertex (e1) at (2,0.75) {$e^-$};
    \vertex (e2) at (2,-0.75) {$e^+$};

    \diagram*{
        (N1L) -- [double, thick] (N1R),
        (N2L) -- [double, thick] (N2R),
        (v1) -- [photon] (v1bis),
        (v2) -- [photon] (v2bis),
        (v1bis) -- [fermion] (e1),
        (v1bis) -- [anti fermion] (v2bis),
        (v2bis) -- [anti fermion] (e2)
    };
\end{feynman}
\end{tikzpicture}
}

\caption{\justifying Feynman diagrams for light-by-light scattering (a), the production of a hadronic resonance decaying into two photons (b), and electron-positron pair production (c) in UPCs.}
\label{fig:background_diagrams}
\end{figure}

The dominant backgrounds for this search arise from light-by-light (LbL) scattering, the production of hadronic resonances decaying into two photons, and $\gamma\gamma \to e^+ e^-$ events in which the electron and positron are misidentified as photons, as shown in Fig.~\ref{fig:background_diagrams}.
The latter two processes are generated directly using \texttt{STARlight}.
For the LbL contribution, we extended the code to include the full leading-order (one-loop) cross section\footnote{Two-loop QCD corrections are negligible within the pseudorapidity range relevant for this analysis~\cite{Bern:2001dg}.}.
We generated $10^7$ events in the center-of-mass energy range $0.1~\mathrm{GeV} \leq \sqrt{s} \leq 6~\mathrm{GeV}$.
For $\sqrt{s} \leq 1~\mathrm{GeV}$, the amplitudes include box diagrams with leptons, pions, and kaons circulating in the loop. At higher energies, $\sqrt{s} > 1~\mathrm{GeV}$, the pion and kaon contributions are replaced by quark box diagrams, following the prescription of Ref.~\cite{Gounaris:1999gh}. We use the $\overline{\rm MS}$  scheme masses $m_e = 5\times 10^{-4}\, {\rm{GeV}}$, $m_{\mu} = 0.106\, {\rm{GeV}}$, $m_{\tau} = 1.777\, {\rm{GeV}}$, $m_u = 2.0\times10^{-3}\, {\rm{GeV}}$, $m_d = 4.7\times10^{-3}$~GeV, $m_s = 9.4\times10^{-2}\, {\rm{GeV}}$ (where the three light-quark masses are evaluated at the renormalization scale $\mu=2$~GeV), $m_c = 1.27\, {\rm{GeV}}$, $m_b = 4.18\, {\rm{GeV}}$, and $m_t = 172.56\, {\rm{GeV}}$ ~\cite{ParticleDataGroup:2024cfk}.  This calculation should be treated with caution across part of the considered energy interval, since the hadronic contribution is not fully captured by the quark-box approximation. In the center-of-mass frame, the final-state photons are predominantly emitted at small angles relative to the incoming photons. Due to the strong collinearity of the initial photons with the ion beams, this background is significantly more forward-peaked than the signal, as illustrated in Fig.~\ref{fig:eta_distr}. Consequently, the pseudorapidity cut based on the detector acceptance effectively suppresses it.

Concerning the production of hadronic resonances, the relevant ones are reported in Table~\ref{tab:hadronic_resonances} along with their quantum numbers, width, production cross section and branching ratio in two photons. For each of them we generated a statistical sample of $10^7$ events. After accounting for the EMC energy resolution and imposing the cut on the photon transverse momentum, the relevant background is essentially due to the decay in two photons of the $\eta^{\prime}$ meson. 
\begin{table*}[t]
    \centering
    \begin{tabular}{ c | c | c | c | c | c | c}
        $X\, (J^P)$ & $m$ (MeV) & $\Gamma$ (MeV) & $\sigma(\gamma\gamma \to X)\, (\mu b)$ & ${\rm Br}(X \to \gamma\gamma)\; (\%)$\\
         \hline
        $\eta\, (0^-)$ & $547.862\pm 0.017$ & $(1.31\pm0.05)\times 10^{-3}$ & $\simeq 980$ & $39.36\pm0.18$\\
        $\eta^{\prime}\, (0^-)$ & $957.78\pm 0.06$ & $0.188 \pm 0.006$ & $\simeq 644$ & $2.307\pm 0.033$\\
        $\eta_c\, (0^-)$ & $2984.1\pm 0.4$ & $30.5\pm 0.5$ & $\simeq 1.5$ & $(1.66\pm 0.13)\times 10^{-2}$\\
    \end{tabular}
    \caption{Hadronic resonances constituting the main background due to their decay in two photons. The table shows their production cross section and branching ratio in two photons.}
    \label{tab:hadronic_resonances}
\end{table*}

Electron--positron pair production proceeds at tree level, and we generated a statistical sample of $10^8$ events.
The pseudorapidity distributions of the final-state electrons and positrons are shown in Fig.~\ref{fig:eta_distr}, illustrating how the detector acceptance efficiently suppresses this background. The misidentification rate of charged particles as photons in the PHENIX detector is expected to be negligible for energies $E > 0.3~\mathrm{GeV}$.
In the absence of a direct measurement, we assume a $1\%$ probability for electrons to fail track reconstruction or track–cluster matching and thus be misidentified as photons\footnote{This is consistent with typical PHENIX central-arm tracking inefficiencies after standard quality cuts in low-multiplicity events~\cite{PHENIX:2002wmm,PHENIX:2003wcn}.}.

Another potential background arises from the production of a pair of neutral pions, each decaying into two photons, where only two of the four photons are detected due to
finite detector efficiency. This background can compete with the other processes discussed above in the region around
$\sqrt{s}\sim 1\,\mathrm{GeV}$, where production is dominated by hadronic resonances~\cite{Klusek-Gawenda:2013rtu}.
By considering all hadronic resonances for which the decay widths into two photons and two neutral
pions have been measured, and assuming a photon detection efficiency of $\sim 70\%$, we
find that the contribution from the $f_2(1270)$ resonance is the only one potentially
comparable to the other backgrounds~\cite{ParticleDataGroup:2024cfk}.
However, once the finite energy resolution of the electromagnetic calorimeter is taken
into account, the other background processes dominate for all values of $\sqrt{s}$.

Finally, we consider the background contribution arising from photon-pomeron interactions, and in particular the process $\gamma + \mathbb{P} \rightarrow J/\psi \rightarrow \eta_c + \gamma$, where the radiative photon from the $J/\psi$ decay is soft enough to escape detection. In this case, the $\eta_c$ subsequently decays into two photons, producing a final state that is experimentally indistinguishable from direct $\eta_c$ photoproduction. The branching ratio for $J/\psi \rightarrow \eta_c + \gamma$ is approximately $1.41\%$~\cite{ParticleDataGroup:2024cfk}, and the radiative photon carries an energy of order $(M^2_{J/\psi} - M^2_{\eta_c}) / (2M_{J/\psi}) \approx 110~\text{MeV}$ in the $J/\psi$ rest frame, which is indeed soft enough to fall below the calorimeter threshold.

\begin{figure}
    \centering
    \includegraphics[width=1.0\linewidth]{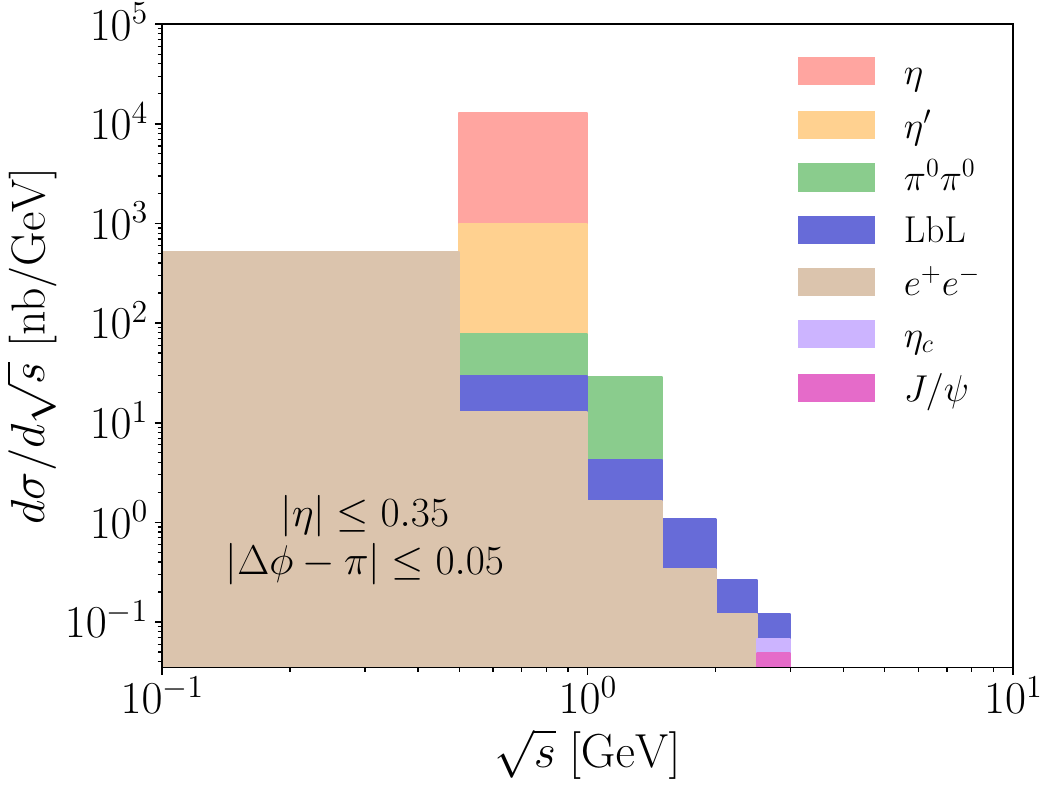}
    \caption{Background cross sections as a function of the two photons invariant mass. The data are shown in energy bins of $0.5~\mathrm{GeV}$ width.}
    \label{fig:background}
\end{figure}

\begin{figure}
    \centering
    \includegraphics[width=1.0\linewidth]{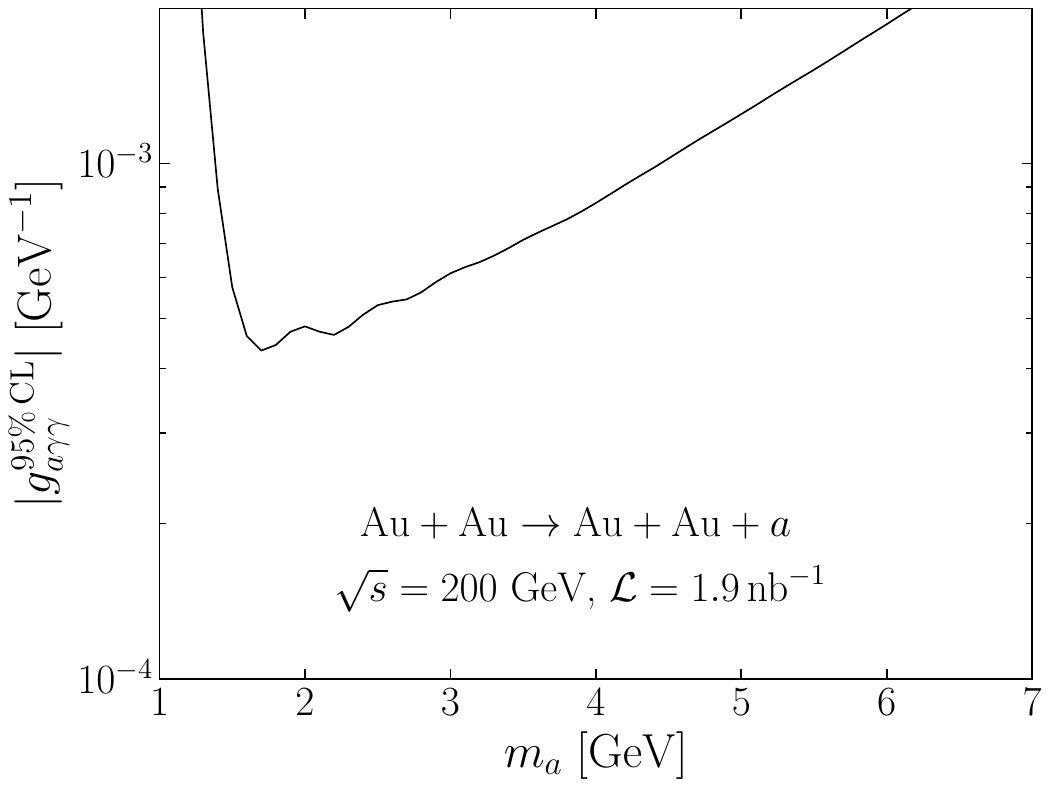}
    \caption{Limits on the axion-photon interaction strength $g_{a\gamma\gamma}$, defined in Eq.~\ref{eq:Lagrangian}, at 95\% C.L. BR($a\rightarrow \gamma\gamma)=1$ is assumed.}
    \label{fig:limits}
\end{figure}

\begin{figure*}
    \centering
    \includegraphics[width=0.9\linewidth]{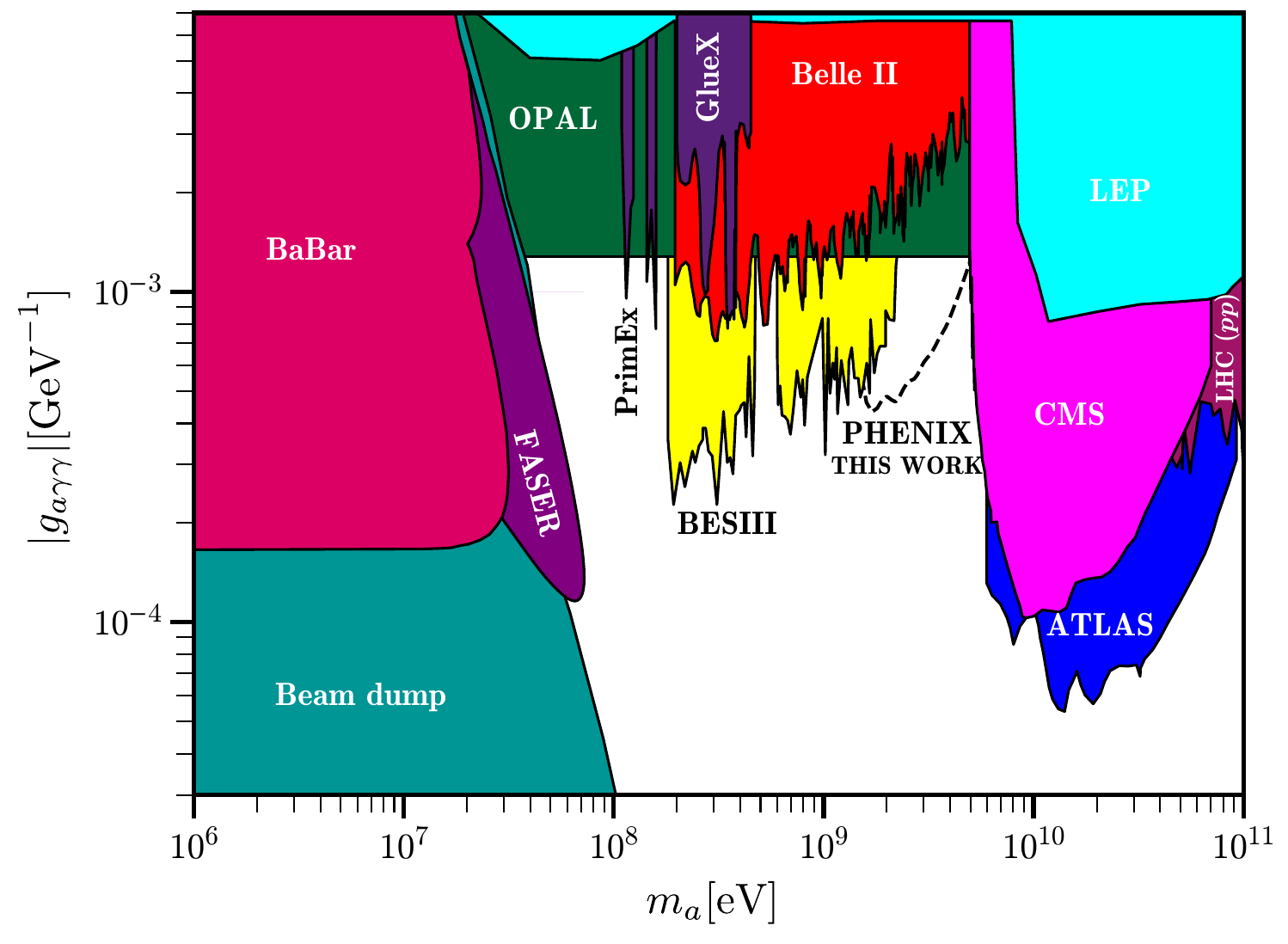}
    \caption{Limits at $95\%$ C.L. on the axion-photon interaction strength computed in this work along with existing constraints~\cite{Dolan:2017osp,CHARM:1985anb,Riordan:1987aw,Dolan:2017osp,Blumlein:1990ay,NA64:2020qwq,FASER:2024bbl,Knapen:2016moh,PrimEx:2010fvg,Aloni:2019ruo,Pybus:2023yex,BESIII:2022rzz,BESIII:2024hdv,Belle-II:2020jti,Jaeckel:2015jla,CMS:2018erd,ATLAS:2020hii}. The PHENIX prospects have been obtained assuming $\mathcal{L} = 1.9\, {\rm nb}^{-1}$ (dashed line). BR($a\rightarrow \gamma\gamma)=1$ is assumed.}
    \label{fig:limits_plot}
\end{figure*}

Figure~\ref{fig:background} shows the background cross sections as a function of the di-photon invariant mass, after applying only the cuts on $\eta$ and $\Delta\phi$ and assuming a perfect energy reconstruction.
The data are binned using a uniform bin width of $0.5\,\mathrm{GeV}$.
Comparison with Fig.~\ref{fig:signal} shows that for $m_a \lesssim 1~\mathrm{GeV}$, the signal is dominated by the background, so that the limits on $g_{a\gamma\gamma}$ are expected to be weak or absent. Hence, we do not gain sensitivity by lowering the threshold on $E_\gamma$. For $\sqrt{s} \gtrsim 1~\mathrm{GeV}$, the background is largely due to the $\eta^\prime$ and LbL scattering, with cross sections comparable to or smaller than the expected signal. Fig.~\ref{fig:events} shows the expected number of events as a function of the di-photon invariant mass for a total integrated luminosity of $1.9~\mathrm{nb}^{-1}$, and $g_{a\gamma\gamma} = 1\, {\rm TeV}^{-1}$, after imposing all the aforementioned cuts and accounting for the EMC energy resolution.\\
Since no dedicated analysis of this dataset has been performed, we derived projected exclusion limits by assuming the data are consistent with the background-only hypothesis. Upper limits on the signal strength are derived at 95\% confidence level using the asymptotic profile-likelihood formalism. For each signal hypothesis, we compute the Asimov test statistic
\begin{equation}
q_\mu = 2 \sum_i \left[ \mu s_i - b_i \ln\!\left(1 + \frac{\mu s_i}{b_i}\right) \right],
\end{equation}
where $s_i$ and $b_i$ denote the expected signal and background in each analysis bin, with the appropriate limiting expression applied for vanishing background. The 95\% CL upper bound on $\mu$ is obtained by numerically solving $q_\mu = 2.71$, corresponding to a one-sided $\chi^2$ criterion. The resulting limits are shown in Fig.~\ref{fig:limits} for a UPC integrated luminosity of $\mathcal{L} = 1.9\, {\rm nb}^{-1}$ as reported by the PHENIX experiment~\cite{PHENIX:2009xtn, Takahara:2012upc}. In Fig.~\ref{fig:limits_plot}, the same limits are shown in the $g_{a\gamma\gamma}$--$m_a$ plane, together with existing experimental constraints in the relevant mass range.

\section{Conclusions}
\label{sec:conclusions}

In this Letter, we have assessed the sensitivity of the PHENIX experiment to ALPs coupled to photons in ultra-peripheral Au-Au collisions at $\sqrt{s} = 200\,\mathrm{GeV}$. The PHENIX detector at RHIC, collected data at a lower collision energy and with a relatively low instantaneous luminosity than heavy-ion experiments at the LHC. These conditions allow photons to be reconstructed and reliably identified down to energies as low as $E_{\gamma} \sim 0.3~\mathrm{GeV}$. In contrast, the higher energies, pileup and trigger limitations at the LHC require typical photon energy thresholds of $\sim 2.5$ GeV. This difference enables PHENIX to probe axion-like particles in a lower-mass region of parameter space that is not accessible in current LHC heavy-ion searches. Exploiting the characteristic $Z^4$ enhancement of the photon-photon luminosity in heavy-ion UPCs, we demonstrate that PHENIX can probe ALPs in the mass range $1\,\mathrm{GeV} \lesssim m_a \lesssim 10\,\mathrm{GeV}$ and with couplings $g_{a\gamma\gamma} \gtrsim 4 \times 10^{-4}\,\mathrm{GeV^{-1}}$ via the resonant process $\gamma\gamma \to a \to \gamma\gamma$. Signal and background processes were simulated using the \textsc{STARlight} Monte Carlo framework, extended to include the full leading-order light-by-light scattering amplitude, and incorporating realistic detector acceptance and energy resolution. The dominant backgrounds, light-by-light scattering, hadronic resonance production
($\eta$, $\eta'$, $\eta_c$), and misidentified $e^+e^-$ pairs, are efficiently suppressed by the detector acceptance and by selection requirements on the photon energies. Assuming an integrated luminosity of $1.9\, {\rm nb}^{-1}$, we derive 95\% confidence level upper limits on the ALP-photon coupling $g_{a\gamma\gamma}$. As shown in Figs.~\ref{fig:limits} and~\ref{fig:limits_plot}, PHENIX is sensitive to previously unexplored regions of the heavy-ALP parameter space, providing a complementary probe to existing constraints from laboratory experiments, LHC measurements, and astrophysical observations. These results underscore the unique potential of the RHIC heavy-ion program for searches of physics beyond the SM. Furthermore, the projected sensitivity estimates motivate a dedicated reanalysis of the existing PHENIX dataset, as well as an assessment of the UPC data collected by other experiments at RHIC, that may have accumulated substantially larger integrated luminosities in Au+Au collisions and may extend the reach of this search further into the ALP parameter space.

\section*{Acknowledgements}
We thank Ming Liu, Spencer Klein, and Barbara Jacak for valuable discussions. K.F. is grateful to N3AS and RIKEN-Berkeley Center at UC Berkeley for hospitality and support, and also acknowledges the support by RIKEN iTHEMS SUURI-COOL (KEK) program.
The work of C.A.M. was supported by the U.S. Department of Energy (DE-SC0009988) and the Sivian Fund. 
The work of H.M. was supported by the NSF grant PHY-2515115, by the U.S. Department of Energy (DE-AC02-05CH11231), by the JSPS Grant-in-Aid for Scientific Research JP23K03382, MEXT Grant-in-Aid for Trans-
formative Research Areas (A) 26H00401, 26A204, 26H00403, Hamamatsu Photonics, K.K, Tokyo Dome Corporation, and by the World Premier International Research Center Initiative (WPI) MEXT, Japan.

\bibliography{biblio}

\end{document}